\documentclass[12pt]{article}
\usepackage{a4wide,graphics,graphicx,amsmath,amssymb,cite,nicefrac}
\usepackage[verbose]{wrapfig}
\usepackage{authblk,hyperref}
\usepackage{url,amsfonts}

\catcode`@=11 \@addtoreset{equation}{section} \catcode`@=12

\begin{document}

\begin{titlepage}
\begin{center}
\hfill DFPD/2017/TH/02\\

\vskip 1.0cm

{\bf \huge Supersymmetric Black Holes\\ \vskip 5pt and Freudenthal Duality}

\vskip 2.0cm

{\bf \large Taniya Mandal${}^{1,2}$,  Alessio Marrani${}^{3,4}$ and Prasanta K. Tripathy${}^{1}$}

\vskip 30pt

{\it ${}^1$%
Department of Physics, Indian Institute of
Technology
Madras,  \\ Chennai
600036, India}\\ \vskip 5pt
{\it ${}^2$%
Institute of Mathematical Sciences,
C.I.T Campus, Taramani\\
Chennai 600113, India}\\ \vskip 5pt
{\it ${}^3$Museo
	Storico
	della Fisica e Centro Studi e Ricerche ``Enrico Fermi'',\\
	Via
	Panisperna
	89A, I-00184, Roma, Italy}\\\vskip 5pt
{\it ${}^4$Dipartimento
	di Fisica e
	Astronomia ``Galileo Galilei'', Universit\`{a} di Padova,\\
	and
	INFN, Sez. di
	Padova,\\Via Marzolo 8, I-35131 Padova, Italy}\\
\vskip 10pt
\texttt{%
	taniya@physics.iitm.ac.in},
\texttt{%
alessio.marrani@pd.infn.it},
\texttt{%
prasanta@iitm.ac.in}

\end{%
center}

\vskip 1.35cm

\begin{%
center} {\bf ABSTRACT}\\[3ex]\end{center}%

We study the effect of
Freudenthal duality on supersymmetric extremal
black
hole attractors in $%
\mathcal{N}=2$, $D=4$ ungauged supergravity.
Freudenthal
duality acts on
the dyonic black hole charges as an
anti-involution which
keeps the black
hole entropy and the critical points
of the effective black
hole potential
invariant. We analyze its effect on
the recently discovered
distinct,
mutually exclusive phases of axionic
supersymmetric black holes,
related
to the existence of non-trivial
involutory constant matrices.
In
particular, we consider a supersymmetric $%
D0-D4-D6$ black hole and
we
explicitly Freudenthal-map it to a
supersymmetric $D0-D2-D4-D6$ black
hole.
We thus show that the charge
representation space of a
supersymmetric $%
D0-D2-D4-D6$ black hole also
contains mutually exclusive
domains.




%
\vfill


\end{titlepage}

\newpage \setcounter{page}{1} \numberwithin{equation}{section}

\section{\label{Intro}Introduction}

Black holes always remain a topic of great interest because of their
thermodynamic behavior. It is one of the significant success of string
theory to provide statistical description (by counting degeneracy of $D$%
-branes in the weak coupling limit) of the macroscopic (Bekenstein-Hawking)
entropy \cite{BH} of certain supersymmetric black holes.

A crucial phenomenon characterizing certain classes of black holes is the
\textit{attractor mechanism}, which states that, for an extremal black hole,
scalar fields are drawn towards fixed values at the horizon losing all
asymptotic memories; such fixed values depend only on black hole electric
and magnetic charges \cite{AM}. A static, spherically symmetric and
asymptotically flat black hole can be described by an effective
one-dimensional theory. The attractors are then obtained by extremizing the
effective black hole potential, that is an algebraic function of the moduli
fields and of the charges. The horizon value of the effective potential
provides the black hole entropy, which depends only on the black hole
charges \cite{FGK}.

An interesting issue in black hole physics concerns the multiplicity of the
attractors within the moduli space; this was first investigated by Moore
\cite{Moore:1998pn}. In $\mathcal{N}=2$, $D=5$ ungauged supergravity,
existence of multiple supersymmetric ($\frac{1}{2}$-BPS) attractors has been
discussed and explicitly constructed for a two-moduli, non-homogeneous model
\cite{Kallosh:1999mz}. In this case, multiplicity arises due to the presence
of disjoint branches in the moduli space \cite{Kallosh:1999mb} (thus being
consistent with the uniqueness results of \cite{Win}). Using the
correspondence between $4D/5D$ critical points \cite{Ceresole:2007rq}, the
analysis has been reduced to four dimensions and it has been shown that a
five-dimensional multiple supersymmetric attractor leads to one
supersymmetric and one non-supersymmetric critical point in four dimensions
\cite{Dominic:2014zia}. While in $D=4$ the supersymmetric attractor is
proved to be unique \cite{Manda:2015zoa}, there exist multiple
non-supersymmetric attractors with the same charge configurations, the same
entropy and the same number of zero modes to the mass matrix (massless
Hessian modes);\ it is puzzling to note that such multiple solutions exist
also when the moduli space is connected \cite{Dominic:2014zia}. By
introducing particular involutory constant matrices (generally depending on
the geometry of the moduli space), in \cite{Manda:2015zoa} it has been also
shown that the representation space of e.m. charges of four dimensional
supersymmetric black holes with non-vanishing axions contains mutually
exclusive domains, and inside each domain the attractor is unique. Within
the same framework, new phases of non-BPS attractors were recently
discovered in \cite{Tripathy-nBPS}.

\bigskip

As the black hole charges transform linearly under symplectic
transformations of the (electric-magnetic) $U$-duality group\footnote{%
Throughout the present investigation, we work in the (semi)classical regime
for which the electromagnetic charges take values in the real numbers. Here $%
U$-duality is referred to as the \textquotedblleft continuous" symmetries of
\cite{Cremmer:1979up}. Their discrete versions are the non-perturbative $U$%
-duality string theory symmetries studied in \cite{Hull:1994ys}.}, the
entropy is a $U$-duality invariant quantity. Four dimensional extremal black
holes also exhibit another transformation which keeps the entropy invariant,
called \textit{Freudenthal duality} (F-duality). F-duality can be defined as
an anti-involutive, non-linear map acting on symplectic spaces, in
particular on the representation space of the black hole electric-magnetic
charges (which becomes a charge lattice upon quantization). After its
introduction \cite{Borsten}, in \cite{FMY-F} F-duality was shown to be a
symmetry not only of the classical Bekenstein-Hawking entropy, but also of
the critical points (attractors) of the effective black hole potential
itself (see also \cite{DualitiesNH, Marrani}). Moreover, it was extended to any generalized special geometry, thus
comprising all $\mathcal{N}>2$ (extended) supergravities, as well as $%
\mathcal{N}=2$ generic, not necessarily homogeneous special K\"{a}hler
geometry \cite{FMY-F}. It is here worth remarking that F-duality appeared in
various other contexts, such as gauge theories with symplectic scalar
manifolds \cite{FG} and multi-centered black holes \cite{FM}. Lagrangian
densities with on-shell F-duality symmetry were constructed in \cite%
{L-F-Dual}. Finally, in \cite{Klemm} F-duality was consistently formulated
in the context of Abelian gaugings of $\mathcal{N}=2$, $D=4$ supergravity,
both for $U(1)$ Fayet-Iliopoulos gauging and for theories coupled to
hypermultiplets.

Despite preserving homogeneity in charges, F-duality is an intrinsically
non-linear, anti-involutive map acting on charges, thus it is inherently
different from $U$-duality, and issues of higher derivative corrections are
not yet well understood. By denoting with $n+1$ the number of Abelian vector
fields, F-duality acts on the dyonic charge vector $Q^{M}=(p^{\Lambda
},q_{\Lambda })$ ($\Lambda =0,1,...,n$) as an entropy-preserving
anti-involution \cite{FMY-F}
\begin{eqnarray}
\pi \hat{Q}^{M}(Q) &:&=\Omega ^{MN}\frac{\partial S(Q)}{\partial Q^{N}},
\label{eq:Fdual} \\
\widehat{\hat{Q}} &=&-Q, \\
S(Q) &=&S(\hat{Q}),
\end{eqnarray}%
where $\Omega $ is the $2(n+1)\times 2(n+1)$ symplectic metric: $\Omega
^{MN}=%
\begin{pmatrix}
0 & -I \\
I & 0%
\end{pmatrix}%
$, $\Omega ^{T}=-\Omega $ and $\Omega ^{2}=-\mathbb{I}$, and $S$ is the
Bekenstein-Hawking (semi)classical black hole entropy. F-duality generally
commutes with supersymmetry (if any) and with $U$-duality; note that no
assumption on the geometry of the moduli space of the scalar fields has been
made.\bigskip

Within the context of cubic geometries of $\mathcal{N}=2$, $D=4$ ungauged
supergravity coupled to $n$ vector multiplets (usually arising from the
large volume limit of compactifications of Type IIA superstrings on
Calabi-Yau manifolds), the aim of this paper is to study the effect of
F-duality on the the most general, axionic $D0-D4-D6$ supersymmetric black
hole solution, which was obtained in \cite{Manda:2015zoa} by introducing
suitable constant, involutory matrices. Considering a simple (yet,
non-homogeneous) $n=2$ model, it was shown that the symplectic
representation space of dyonic black hole charges (which becomes a charge
lattice upon imposing Dirac-Schwinger-Zwanzinger quantization conditions)
contains mutually exclusive domains, and that inside each domain the
attractor solution is unique \cite{Manda:2015zoa}. By F-transforming the
most general $D0-D4-D6$ BPS black hole attractors, we will prove that such
mutually exclusive domains also exist for the most general $D0-D2-D4-D6$ BPS
supporting charge configuration. Moreover, we will formulate the most
general \textit{Ansatz} for the entropy and the value of the horizon,
attracted moduli within such a configuration, thus fully generalizing the
results of \cite{Shmakova:1996nz} and \cite{Manda:2015zoa}.\bigskip

The plan of the paper is as follows.

In Sec. \ref{solutions}, we recap the $D0-D4-D6$ BPS black hole attractor
solution, while in Sec. \ref{Sec-n=2} we consider a model with two moduli
and state the condition of existence of a non-trivial involutory matrix. We
further comment on the non-homogeneity and explicit realization of such a
model, as well as on the $U$-invariance, in Sec. \ref{Comments}.

Then, in Sec. \ref{sec3} we apply F-duality on the $D0-D4-D6$ BPS black
hole, finding that the corresponding F-dual configuration is a $D0-D2-D4-D6$
BPS black hole.

Sec. \ref{sec4} is devoted to the generalization of the results of \cite%
{Shmakova:1996nz} and \cite{Manda:2015zoa} : we discuss the most general
supersymmetric $D0-D2-D4-D6$ black hole attractor solution which, for
particular values of real, charge-dependent parameters, is F-dual to $%
D0-D4-D6$.

Sec. \ref{sec5} presents some comments and hints for further developments.

An Appendix, containing some technical details, concludes the paper.

\section{ \label{sec2}Supersymmetric $D0-D4-D6$ Black Hole}

\subsection{\label{solutions}The Solutions}

Attractor solutions describing supersymmetric extremal $D0-D4-D6$ black
holes in $\mathcal{N}=2$, $D=4$ supergravity have been studied in \cite%
{Manda:2015zoa}. In a special K\"{a}hler geometry of the vector multiplets'
moduli space described by the holomorphic prepotential ($a=1,...,n$)%
\begin{equation}
F(X)=D_{abc}\frac{X^{a}X^{b}X^{c}}{X^{0}},  \label{hol-F}
\end{equation}%
the most general supersymmetric attractor solution to this black hole
configuration is given by
\begin{eqnarray}
x_{1}^{a} &=&\frac{1}{p^{0}}\left( p^{a}-\frac{2D-q_{0}\left( {p^{0}}\right)
^{2}}{2\chi }{I^{a}}_{b}p^{b}\right) ,  \label{eq:mostgensol} \\
x_{2}^{a} &=& - \frac{1}{p^{0}}\sqrt{1-\left( \frac{2D-q_{0}\left( {p^{0}}%
\right) ^{2}}{2\chi }\right) ^{2}}{I^{a}}_{b}p^{b},  \label{eq:mostgensol-2}
\end{eqnarray}%
where the so-called $4D/5D$ special coordinates' symplectic frame is used%
\footnote{%
In such a symplectic frame, the $D$-brane charges can be denoted as follows
: $D0$ ($q_{0}$), $D2$ ($q_{a}$), $D4$ ($p^{a}$) and $D6$ ($p^{0}$), as
understood throughout.}, with the scalar fields denoted by $x^{a}:=$ $%
x_{1}^{a}+ix_{2}^{a}$. Moreover, we used the notation of \cite{Manda:2015zoa}%
, namely $D:=D_{abc}p^{a}p^{b}p^{c}$, $D_{a}:=D_{abc}p^{b}p^{c}$ and $\chi
:=D_{a}{I^{a}}_{b}p^{b}$. The involutory matrix ${I^{a}}_{b}$ satisfies \cite%
{Manda:2015zoa}
\begin{eqnarray}
{I^{a}}_{c}{I^{c}}_{b} &=&{\delta ^{a}}_{b}  \label{eq:involution} \\
D_{ade}{I^{d}}_{b}{I^{e}}_{c} &=&D_{abc}.  \label{eq:involution-2}
\end{eqnarray}%
The corresponding black hole entropy takes the following form \cite%
{Manda:2015zoa}:
\begin{equation}
S=\frac{\pi }{p^{0}}\sqrt{4\chi ^{2}-(2D-q_{0}{p^{0}}^{2})^{2}}.
\label{eq:mostgen}
\end{equation}%
Any choice of ${I^{a}}_{b}$ that satisfy conditions (\ref{eq:involution})-(%
\ref{eq:involution-2}) will provide an independent and well-defined
expression of the Bekenstein-Hawking entropy and of corresponding attractor,
horizon values of the scalar fields (\ref{eq:mostgensol})-(\ref%
{eq:mostgensol-2}). The trivial choice of the involution is of course ${I^{a}%
}_{b}={\delta ^{a}}_{b}$. For this choice, (\ref{eq:mostgen}) reduces to
\begin{equation}
S=\pi \sqrt{4q_{0}D-(q_{0}p^{0})^{2}},  \label{eq:standardS}
\end{equation}%
which is the standard entropy of $D0-D4-D6$ BPS extremal black hole \cite%
{Shmakova:1996nz}. The attractor value of the scalar fields for this trivial
choice of involutory matrix is\footnote{%
From the treatment of \cite{Ceresole:2007rq}, Im$\left( x^{a}\right) <0$ in
the $4D/5D$ special coordinates' symplectic frame of special K\"{a}hler
geometry.} \cite{Shmakova:1996nz, Manda:2015zoa}
\begin{equation}
x^{a}=\frac{p^{a}}{2D}\left( p^{0}q_{0}-i\sqrt{4q_{0}D-(p^{0}q_{0})^{2}}%
\right) .  \label{eq:standard}
\end{equation}

Note that the above expression reduces to the smooth $D0-D4$ attractor
solution upon setting the $D6$ charge $p^0$ to zero. However, for any
other choice of ${I^a}_b$ the solutions \eqref{eq:mostgensol} and
\eqref{eq:mostgensol-2} as well as the corresponding entropy \eqref{eq:mostgen}
become singular in the limit $p^0\rightarrow 0$.
Thus, these new branches of solutions cease to exist in the absence of $D6$
branes. This is consistent with the uniqueness of $D0-D4$ attractors
\cite{Manda:2015zoa}.

\subsection{\label{Sec-n=2}$n=2$ Model and Non-trivial Involutory Matrix}

By considering an example of two-moduli ($n=2$) model, in Sec. 4 of \cite%
{Manda:2015zoa}, it has also been shown explicitly that there exists a
non-trivial choice of an involutory matrix ${I^{a}}_{b}\neq {\delta ^{a}}%
_{b} $ satisfying (\ref{eq:involution})-(\ref{eq:involution-2}). A generic $%
2\times 2$ involutory matrix can be parametrized as
\begin{equation}
{I^{a}}_{b}=:%
\begin{pmatrix}
u & v \\
w & -u%
\end{pmatrix}%
,  \label{eq:nontrivinvolution}
\end{equation}%
such that $u^{2}+vw=1$; for brevity's sake, we further denote the four
possible components of the rank-3 completely symmetric tensor $D_{abc}$
occurring in (\ref{hol-F}) (namely, of the triple intersection number of the
Calabi-Yau manifold in the large volume limit of the Type IIA
compactifications) as $D_{111}=a$, $D_{112}=b$, $D_{122}=c$ and $D_{222}=d$.
It has been proved that the conditions (\ref{eq:involution})-(\ref%
{eq:involution-2}) can be satisfied with ${I^{a}}_{b}\neq {\delta ^{a}}_{b}$%
, for%
\begin{equation}
u=\frac{\mathcal{L}}{\sqrt{{\mathcal{L}}^{2}-4\mathcal{MN}}},~v=\frac{-2%
\mathcal{M}}{\sqrt{{\mathcal{L}}^{2}-4\mathcal{MN}}},~w=\frac{2\mathcal{N}}{%
\sqrt{{\mathcal{L}}^{2}-4\mathcal{MN}}},  \label{eq:nontrivinvolution-2}
\end{equation}%
where we introduced the notation $\mathcal{L}=ad-bc,\mathcal{M}=c^{2}-bd$
and $\mathcal{N}=b^{2}-ac$. Note that a crucial consistency condition is%
\begin{equation}
{\mathcal{L}}^{2}-4\mathcal{MN}>0.  \label{consistency-cond}
\end{equation}%
For the choice (\ref{eq:nontrivinvolution}-(\ref{eq:nontrivinvolution-2}) of
${I^{a}}_{b}$, the horizon, attractor values of the two vector multiplet
moduli $x^{1}=x_{1}^{1}+ix_{2}^{1}$ and $x^{2}=x_{1}^{2}+ix_{2}^{2}$ are
given by (\ref{eq:mostgensol})-(\ref{eq:mostgensol-2}), which explicitly read%
\begin{equation}
x_{1}^{1}=\frac{1}{p^{0}}\left[ p^{1}-\frac{\left( D-\frac{1}{2}q_{0}\left( {%
p^{0}}\right) ^{2}\right) (\mathcal{L}p^{1}-2\mathcal{M}p^{2})}{\chi \sqrt{{%
\mathcal{L}}^{2}-4\mathcal{M}\mathcal{N}}}\right] ,  \label{sol-1}
\end{equation}%
\begin{equation}
x_{2}^{1}= - \frac{1}{p^{0}}\sqrt{1-\left( {\frac{D-\frac{1}{2}q_{0}\left( {%
p^{0}}\right) ^{2}}{\chi }}\right) ^{2}}\frac{(\mathcal{L}p^{1}-2\mathcal{M}%
p^{2})}{\sqrt{{\mathcal{L}}^{2}-4\mathcal{M}\mathcal{N}}},  \label{sol-2}
\end{equation}%
\begin{equation}
x_{1}^{2}= \frac{1}{p^{0}}\left[ p^{2}-\frac{\left( D-\frac{1}{2}%
q_{0}\left( {p^{0}}\right) ^{2}\right) (2\mathcal{N}p^{1}-\mathcal{L}p^{2})}{%
\chi \sqrt{{\mathcal{L}}^{2}-4\mathcal{M}\mathcal{N}}}\right] ,
\label{sol-3}
\end{equation}%
\begin{equation}
x_{2}^{2}= - \frac{1}{p^{0}}\sqrt{1-\left( {\frac{D-\frac{1}{2}q_{0}\left( {%
p^{0}}\right) ^{2}}{\chi }}\right) ^{2}}\frac{(2\mathcal{N}p^{1}-\mathcal{L}%
p^{2})}{\sqrt{{\mathcal{L}}^{2}-4\mathcal{M}\mathcal{N}}}.  \label{sol-4}
\end{equation}%
So, for an $n=2$ model satisfying (\ref{consistency-cond}), two distinct and
independent BPS attractor solutions exist, depending on whether ${I^{a}}%
_{b}=\delta _{b}^{a}$ or ${I^{a}}_{b}$ given by (\ref{eq:nontrivinvolution}-(%
\ref{eq:nontrivinvolution-2}) is chosen.

Now the key question naturally arises whether these two BPS solutions exist
for the same supporting $D0-D4-D6$ black hole charge configuration. As
investigated in \cite{Manda:2015zoa}, the conditions of positive
definiteness of the K\"{a}hler metric $g_{a\bar{b}}$ for the two classes of
BPS solutions boil down to the study of the sign of the quantity%
\begin{equation}
\mathcal{N}\left( {p^{1}}\right) ^{2}-\mathcal{L}p^{1}p^{2}+\mathcal{M}%
\left( {p^{2}}\right) ^{2},  \label{2-par-quantity}
\end{equation}%
and they read as follows :
\begin{eqnarray}
\text{`std.'~sol.~(\ref{eq:standardS})-(\ref{eq:standard})}\text{:~} &&%
\mathcal{N}\left( {p^{1}}\right) ^{2}-\mathcal{L}p^{1}p^{2}+\mathcal{M}%
\left( {p^{2}}\right) ^{2}>0;  \label{std-1} \\
\text{`non-std.'~sol.~(\ref{eq:mostgen}) and (\ref{sol-1})-(\ref{sol-4})}%
\text{:~} &&\mathcal{N}\left( {p^{1}}\right) ^{2}-\mathcal{L}p^{1}p^{2}+%
\mathcal{M}\left( {p^{2}}\right) ^{2}<0.  \label{std-2}
\end{eqnarray}%
Also the positive definiteness of the gauge kinetic terms can (\textit{at
least} numerically) be checked to hold for both solutions for such distinct
conditions. Thus, in the considered $n=2$ model, the `standard' and
`non-standard' BPS attractor solutions are supported by two different,
distinct domains in the representation space of the dyonic black hole
charges \cite{Manda:2015zoa}.

\subsubsection{\label{Comments}Non-Homogeneity and $U$-Invariance}

Some comments are in order here.

\begin{enumerate}
\item The consistency condition (\ref{consistency-cond}) necessarily implies
the non-homogeneity of the vector multiplets's two-dimensional special K\"{a}%
hler moduli space. Indeed, from the classification of homogeneous ($d$%
-)spaces \cite{dWVP,dWVVP}, the unique $n=2$ homogeneous model is the
so-called $ST^{2}$ model (corresponding to the dimensional reduction of $%
(1,0)$, chiral minimal \textquotedblleft pure" supergravity from $D=6$ to $%
D=4$ \cite{dWVVP}), characterized by $c=2$ and $a=b=d=0$, and thus violating
(\ref{consistency-cond}), because in this case ${\mathcal{L}}^{2}-4\mathcal{%
MN}=0$. Thus, the $n=2$ model considered here, as well as in \cite%
{Manda:2015zoa}, has a \textit{non-homogeneous} moduli space.

\item For explicit $n=2$ models for which the treatment of Sec. \ref{Sec-n=2}
holds, one can \textit{e.g.} see the two Tables at pages 51-53 of \cite%
{alg-geom}, providing various (non-homogeneous) cubic models in which
multiple axionic $D=4$ BPS attractors exist. For instance, $%
X_{14}^{II}\left( 7,2,2,2,1\right) $, for which $a=2$, $b=7$, $c=21$, $d=63$
is an example of such models \cite{alg-geom}.

\item As also resulting from the treatment \textit{e.g.} of \cite{dWVVP},
all special K\"{a}hler geometries described by the cubic holomorphic
prepotential (\ref{hol-F}) (usually named $d$-geometries) are characterized
by a `minimal' electric-magnetic ($U$-)duality symmetry, which pertains to
(the large volume limit of) Calabi-Yau compactifications (\cite{DF}; for
recent accounts, \textit{cfr.} \cite{d-geom-revisited} and \cite%
{Klemm-quantum-STU}, and Refs. therein). It can easily be checked that the
two inequality conditions appearing in the r.h.s.'s of (\ref{std-1}) and (%
\ref{std-2}) are not invariant under such a duality, and thus they are not
well-defined. In fact, in order to $U$-invariantly characterize the
separation of the two domains in the charge representation space, one should
better consider the positivity condition of the determinant of the metric $%
g_{a\bar{b}}$ for both classes of BPS solutions, respectively given by Eqs.
(4.8) and (4.9) of \cite{Manda:2015zoa}, which are consistently $U$%
-invariant.
\end{enumerate}

\section{\label{sec3}Freudenthal Dual of $D0-D4-D6$ BPS Black Hole}

In this section we apply F-duality on the $D0-D4-D6$ charge vector $Q$ and
find a new dyonic charge vector $\hat{Q}$. Following (\ref{eq:Fdual}) and
also the relation between moduli and entropy (\textit{cfr. e.g.} \cite%
{Ferrara:2006yb}, and Refs. therein)
\begin{equation}
x^{a}=\frac{p^{a}+\frac{i}{\pi }\frac{\partial S}{\partial q_{a}}}{p^{0}+%
\frac{i}{\pi }\frac{\partial S}{\partial q_{0}}},
\end{equation}%
we obtain the new F-transformed set of charges $\hat{Q}=\left( \hat{p}^{0},%
\hat{p}^{a},\hat{q}_{0},\hat{q}_{a}\right) $ in terms of $%
Q=(p^{0},p^{a},q_{0},0_{a})$, where $S$ is the most general supersymmetric
entropy \cite{Shmakova:1996nz} :
\begin{eqnarray}
\hat{p}^{0} &=&\frac{\pi (q_{0}{p^{0}}^{2}-2D)}{S},  \label{c1} \\
\hat{p}^{a} &=&\frac{\pi }{p^{0}S}\left[ 2\chi {I^{a}}_{b}p^{b}-\left(
2D-q_{0}\left( {p^{0}}\right) ^{2}\right) p^{a}\right] ,  \label{c2} \\
\hat{q}_{0} &=&-\frac{\pi }{{p^{0}}^{3}S}\left[ 4\chi ^{2} - 4 D^2 +
q_0^2(p^0)^4\right] ,  \label{c3} \\
\hat{q}_{a} &=&\frac{6\pi }{{p^{0}}^{2}S}\left[ 2\chi D_{b}{I^{b}}_{a}-
\left( 2 D - q_{0}\left( {p^{0}}\right) ^{2}\right) D_{a}\right] .
\label{c4}
\end{eqnarray}%
So, we have obtained a generic charge configuration (with all types of $D$%
-brane charges switched on) by acting with F-duality on the supersymmetric $%
D0-D4-D6$ black hole : this is provided by the most general F-dualized
charge configuration $\hat{Q}$ coming from supersymmetric $D0-D4-D6$ black
hole charge configuration $Q$.

\section{\label{sec4}Supersymmetric $D0-D2-D4-D6$ Black Hole}

By further generalizing the results of \cite{Manda:2015zoa}, in this section
we will find out the most general expression of entropy and moduli of a
supersymmetric $D0-D2-D4-D6$ black hole, and show that it is F-dual to the
supersymmetric $D0-D4-D6$ with the help of Eqs. (\ref{c1})-(\ref{c4}). For
convenience's sake, we will denote all generic BPS dyonic $D0-D2-D4-D6$
charges, as well as the symplectic vector comprising them all, as tilded.

In order to find out the entropy and attractor solution, we follow the
method due to Shmakova \cite{Shmakova:1996nz}. For a generic supersymmetric
black hole in Type IIA string theory with all non-vanishing $D$-brane
charges, the entropy is given by \cite{Shmakova:1996nz} :
\begin{equation}
\frac{S(\tilde{Q})}{\pi }=\frac{1}{3\tilde{p}^{0}}\sqrt{\frac{4}{3}(\tilde{%
\Delta}_{a}\tilde{x}^{a})^{2}-9(\tilde{p}^{0}(\tilde{p}\cdot \tilde{q})-2%
\tilde{D})^{2}},  \label{eq:entropy}
\end{equation}%
where $\tilde{\Delta}_{a}:=3\tilde{D}_{a}-\tilde{p}^{0}\tilde{q}_{a}$, $%
\tilde{D}_{a}:=D_{abc}\tilde{p}^{b}\tilde{p}^{c}$, $\tilde{D}:=\tilde{D}_{a}%
\tilde{p}^{a}$ and $\tilde{p}\cdot \tilde{q}:=\tilde{p}^{0}\tilde{q}_{0}+%
\tilde{p}^{a}\tilde{q}_{a}$. The variables $\tilde{x}^{a}$ are the real
solutions of
\begin{equation}
D_{abc}\tilde{x}^{b}\tilde{x}^{c}=\tilde{\Delta}_{a}.  \label{eq:tilde}
\end{equation}%
The moduli fields $x^{a}=x_{1}^{a}+ix_{2}^{a}$ at the BPS attractor point
are \cite{Shmakova:1996nz}
\begin{eqnarray}
x_{1}^{a} &=&\frac{3}{2}\frac{\tilde{x}^{a}}{\tilde{p}^{0}(\tilde{\Delta}_{c}%
\tilde{x}^{c})}(\tilde{p}^{0}(\tilde{p}\cdot \tilde{q})-2\tilde{D})+\frac{%
\tilde{p}^{a}}{\tilde{p}^{0}},  \label{eq:modulig} \\
x_{2}^{a} &=&-\frac{3}{2}\frac{\tilde{x}^{a}}{(\tilde{\Delta}_{c}\tilde{x}%
^{c})}\frac{S(\tilde{Q})}{\pi }.  \label{eq:modulig-2}
\end{eqnarray}%
To solve Eq. (\ref{eq:tilde}), we consider the most general \textit{Ansatz}

\begin{equation}
\tilde{x}^{a}=\alpha \tilde{p}^{0}{I^{a}}_{b}\tilde{D}^{bc}\tilde{q}%
_{c}+\beta \tilde{p}^{0}\tilde{D}^{ab}\tilde{q}_{b}+\sigma {I^{a}}_{b}\tilde{%
p}^{b}+\rho \tilde{p}^{a},  \label{eq:tildeansatz}
\end{equation}%
where $\tilde{D}^{ac}\tilde{D}_{cb}:=\delta _{b}^{a}$, $\tilde{D}%
_{ab}:=D_{abc}\tilde{p}^{c}$, and $\alpha ,\beta ,\sigma $ and $\rho $ are
real, $\tilde{Q}$-dependent quantities. From the very definition of $\tilde{%
\Delta}_{a}$ and Eq. (\ref{eq:tilde}), one obtains
\begin{eqnarray}
\tilde{\Delta}_{a} &=&(\tilde{p}^{0})^{2}(\alpha ^{2}+\beta ^{2})D_{abc}%
\tilde{D}^{be}\tilde{q}_{e}\tilde{D}^{cf}\tilde{q}_{f}+(\sigma ^{2}+\rho
^{2})\tilde{D}_{a}+2\tilde{p}^{0}(\alpha \sigma +\beta \rho )\tilde{q}_{a}
\notag \\
&&+2\tilde{p}^{0}(\alpha \rho +\beta \sigma )\tilde{q}_{b}{I^{b}}_{a}+2(%
\tilde{p}^{0})^{2}\alpha \beta D_{abc}{I^{b}}_{d}\tilde{D}^{de}\tilde{q}_{e}%
\tilde{D}^{cf}\tilde{q}_{f}  \notag \\
&&+2\sigma \rho D_{abc}{I^{b}}_{d}\tilde{p}^{d}\tilde{p}^{c}.
\label{Delta-hat}
\end{eqnarray}%
The most general BPS dyonic solution will be given by the most general
solution of Eq. (\ref{eq:tilde}) with (\ref{Delta-hat}) plugged in.

By recalling the results (\ref{c1})-(\ref{c4}), one can check (\textit{cfr.}
also the Appendix) that setting%
\begin{equation}
\alpha =-\beta =-\frac{\mathbb{B}}{2\sqrt{3}\mathbb{A}}\frac{\sqrt{(6\tilde{D%
}-\tilde{p}^{0}A)^{2}-(\tilde{p}^{0}B)^{2}}}{(6\tilde{D}-\tilde{p}^{0}A)},
\label{4.7}
\end{equation}%
\begin{equation}
\sigma =\frac{\sqrt{3}(6\tilde{D}-\tilde{p}^{0}A)}{\sqrt{(6\tilde{D}-\tilde{p%
}^{0}A)^{2}-(\tilde{p}^{0}B)^{2}}},
\end{equation}%
\begin{equation}
\rho =-\frac{\sqrt{3}\tilde{p}^{0}B}{\sqrt{(6\tilde{D}-\tilde{p}^{0}A)^{2}-(%
\tilde{p}^{0}B)^{2}}},
\end{equation}%
where%
\begin{eqnarray}
A &:&=\tilde{q}_{a}\tilde{p}^{a},~~B:=\tilde{q}_{a}{I^{a}}_{b}\tilde{p}%
^{b},~~ \\
\mathbb{A} &:&=18\tilde{q}_{0}+\tilde{D}^{ab}\tilde{q}_{a}\tilde{q}_{b},~~%
\mathbb{B}:=\tilde{q}_{a}{I^{a}}_{b}\tilde{D}^{bc}\tilde{q}_{c},~~
\label{X-Y}
\end{eqnarray}%
the supersymmetric $D0-D2-D4-D6$ black hole charge vector $\tilde{Q}$
becomes F-dual to the supersymmetric $D0-D4-D6$ black hole charge vector $Q$
:%
\begin{equation}
\left. \tilde{Q}\right\vert _{\text{(\ref{c1})-(\ref{c4}),~(\ref{4.7})-(\ref%
{X-Y})}}=\hat{Q},
\end{equation}%
and all tilded quantities correspondingly become hatted. This duality
trivially holds for the `standard' BPS solution with ${I^{a}}_{b}={\delta
^{a}}_{b}$, providing the BPS solution found by Shmakova \cite%
{Shmakova:1996nz}. On the other hand, for the new, `non-standard' solution
with ${I^{a}}_{b}\neq {\delta _{b}^{a}}$, $\tilde{x}^{a}$ changes, and so $%
x_{1}^{a}$, $x_{2}^{a}$ and $S$ do, and we find distinct solutions.

When all the brane charges are non-vanishing, the charge-dependent
quantities $\mathbb{A}$ and $\mathbb{B}$ defined in (\ref{X-Y}) satisfy the
relation
\begin{equation}
\mathbb{A}(6\hat{D}_{a}-\hat{p}^{0}\hat{q}_{a})+\hat{p}^{0}\mathbb{B}\hat{q}%
_{b}{I^{b}}_{a}=0.  \label{X-Y-rel}
\end{equation}%
Exploiting Eqs. (\ref{c1})-(\ref{c4}), one can then write the $D4$ brane
charges $p^{a}$ of supersymmetric $D0-D4-D6$ black hole in terms of
supersymmetric $D0-D2-D4-D6$ black hole charges as
\begin{equation}
p^{a}=\frac{\alpha \hat{p}^{0}}{\sqrt{3}}\left( \hat{D}^{ab}\hat{q}_{b}-{I^{a}}_{b}%
\hat{D}^{bc}\hat{q}_{c}\right) +\frac{1}{\sqrt{3}}\left( \sigma \hat{p}%
^{a}+\rho {I^{a}}_{b}\hat{p}^{b}\right) .  \label{p-rel}
\end{equation}

For $n=2$ (two-moduli cubic model, treated in Sec. \ref{Sec-n=2}), by
plugging (\ref{p-rel}) into (\ref{2-par-quantity}), one can express this
latter in terms of hatted charges. Thus, as for the $D0-D4-D6$ case, also
for the $D0-D2-D4-D6$ case one can explicitly check that the `standard' and
`non-standard' classes of BPS solutions are supported by different, mutually
exclusive domains in the charge representation space.

Finally, it is worth pointing out that if one sets $\tilde{q}^{a}=0$, $%
\alpha $, $\beta $ and $\rho $ vanish and $\sigma =\sqrt{3}$ and then, by
inserting them in Eq. (\ref{eq:tilde}), we find the most general
supersymmetric $D0-D4-D6$ entropy and attractor solutions in tilded charges
\cite{Manda:2015zoa}. Thus, one can conclude that (\ref{eq:tildeansatz}) is
the most general BPS \textit{Ansatz} in $\mathcal{N}=2$, $D=4$ ungauged
supergravity coupled to vector multiplets, up to the constraint (\ref%
{eq:tilde}). This is the full-fledged generalization of\ \cite%
{Shmakova:1996nz} for completely general BPS dyonic $D0-D2-D4-D6$ supporting
charge configuration.

\section{\label{sec5}Conclusion}

In $\mathcal{N}=2$, $D=4$ ungauged supergravity coupled to vector multiplets
with cubic holomorphic prepotential (\ref{hol-F}), by using suitable
involutory matrices, we have constructed the most general supersymmetric ($%
\frac{1}{2}$-BPS) attractor solution of $D0-D2-D4-D6$ extremal black hole
which is Freudenthal dual to the most general attractor solution of
supersymmetric $D0-D4-D6$ black hole. This holds for suitable choices of the
parameters $\alpha $, $\beta $, $\sigma $ and $\rho $ of the solution.

In an $n=2$ model with non-homogeneous moduli space, the charge
representation space supporting supersymmetric $D0-D2-D4-D6$ extremal black
hole attractors contains mutually exclusive domains, and inside each of
these domains there exists a unique supersymmetric attractor.

Especially in the cases with more than two moduli, it would be interesting
to further explore the geometric constraints on the existence of non-trivial
involutory matrices satisfying (\ref{eq:involution})-(\ref{eq:involution-2}%
), and thus of `non-standard' classes of BPS attractors (which, for $n=2$,
are given by the consistency condition (\ref{consistency-cond})). The known
orbit stratification of the charge representation space in models with
symmetric moduli spaces would prevent the existence of `non-standard'
classes of BPS attractors; it remains to be seen whether this is also the
case for homogeneous non-symmetric moduli spaces, as well. We leave this
very interesting issue for further future investigation.

For what concerns the study of involutory matrices in non-homogeneous
special K\"{a}hler moduli spaces arising in (the large volume limit of) Type
IIA compactifications on Calabi-Yau manifolds, one might consider the
relatively simple class of \textit{reducible} cubic prepotentials, which was
recently classified in \cite{Cortes}, showing that the automorphism group in
the two non-homogeneous classes of such spaces has a co-homogeneity one
action, thus of a peculiar non-transitive type. We hope to report on such
topics in future works.

\newpage

\appendix

\section{Appendix}

We insert $D0-D2-D4-D6$ F-dual charges $\hat{Q}$ in terms of $D0-D4-D6$
charges $Q$ in the expressions made of $\hat{Q}$, using Eqs. (\ref{c1}), (%
\ref{c2}), (\ref{c3}), (\ref{c4}), (\ref{eq:tildeansatz}) and (\ref%
{Delta-hat}). After some algebra, we find
\begin{equation}
\alpha (\hat{Q})=-\beta (\hat{Q})=-\frac{\sqrt{4\chi ^{2}-\left(
2D-q_{0}\left( {p^{0}}\right) ^{2}\right) ^{2}}}{2\sqrt{3}\left(
2D-q_{0}\left( {p^{0}}\right) ^{2}\right) },
\end{equation}%
\begin{equation}
\sigma (\hat{Q})=\frac{2\sqrt{3}\chi }{\sqrt{4\chi ^{2}-\left(
2D-q_{0}\left( {p^{0}}\right) ^{2}\right) ^{2}}},
\end{equation}%
\begin{equation}
\rho (\hat{Q})=\frac{\sqrt{3}\left( 2D-q_{0}\left( {p^{0}}\right)
^{2}\right) }{\sqrt{4\chi ^{2}-\left( 2D-q_{0}\left( {p^{0}}\right)
^{2}\right) ^{2}}},
\end{equation}%
\begin{equation}
\tilde{x}^{a}(\hat{Q})=\sqrt{3}{I^{a}}_{b}p^{b},
\end{equation}%
\begin{equation}
\tilde{\Delta} _{a}\tilde{x}^{a}(\hat{Q})=3\sqrt{3}\chi .
\end{equation}%
As it is well known, the entropy is F-invariant \cite{Borsten, FMY-F} (%
\textit{cfr.} (\ref{eq:entropy})) :
\begin{equation}
\frac{S(\hat{Q})}{\pi }=\frac{S(Q)}{3\pi (q_{0}{p^{0}}^{2}-2D)}\sqrt{\frac{4%
}{3}(3\sqrt{3}\chi )^{2}-9(4\chi ^{2}-(2D-q_{0}{p^{0}}^{2})^{2})}=\frac{S(Q)%
}{\pi },
\end{equation}%
and so are the horizon, attractor values of the moduli \cite{FMY-F} (\textit{%
cfr.} (\ref{eq:modulig})-(\ref{eq:modulig-2}))
\begin{eqnarray}
x_{1}^{a}(\hat{Q}) &=&\frac{3}{2}\frac{{I^{a}}_{b}p^{b}}{3\chi }\left( \frac{%
S}{\pi }\right) ^{2}\frac{p^{0}}{\left( 2D-q_{0}\left( {p^{0}}\right)
^{2}\right) }-\frac{2\chi {I^{a}}_{b}p^{b}-\left( 2D-q_{0}\left( {p^{0}}%
\right) ^{2}\right) p^{a}}{p^{0}\left( 2D-q_{0}\left( {p^{0}}\right)
^{2}\right) },  \notag \\
&=&\frac{1}{p^{0}}\left( p^{a}-\frac{2D-q_{0}\left( {p^{0}}\right) ^{2}}{%
2\chi }{I^{a}}_{b}p^{b}\right) =x_{1}^{a}(Q), \\
x_{2}^{a}(\hat{Q}) &=&-\frac{3}{2}\frac{{I^{a}}_{b}p^{b}}{3\chi }\frac{S(Q)}{%
\pi }=-\frac{1}{p^{0}}\sqrt{1-\left( \frac{2D-q_{0}\left( {p^{0}}\right) ^{2}%
}{2\chi }\right) ^{2}}{I^{a}}_{b}p^{b}=x_{2}^{a}(Q).
\end{eqnarray}


\newpage

\begin{thebibliography}{99}
\bibitem{BH} S. W. Hawking, \textit{Gravitational Radiation from Colliding
Black Holes}, Phys. Rev. Lett. \textbf{26} (1971) 1344. J. D. Bekenstein,
\textit{Black Holes and Entropy}, Phys. Rev. \textbf{D7} (1973) 333.

\bibitem{AM} S. Ferrara, R. Kallosh, A. Strominger, $\mathcal{N}=2$ \textit{%
Extremal Black Holes}, Phys. Rev. \textbf{D52} (1995) 5412, \texttt{%
hep-th/9508072}. A. Strominger, \textit{Macroscopic Entropy of }$\mathcal{N}%
=2$\textit{\ Extremal Black Holes}, Phys. Lett. \textbf{B383}, 39 (1996),
\texttt{hep-th/9602111}. S.~Ferrara, R.~Kallosh, \textit{Supersymmetry and
attractors}, Phys.\ Rev.\ \textbf{D54} (1996) 1514, \texttt{hep-th/9602136}.
S. Ferrara, R. Kallosh, \textit{Universality of Supersymmetric Attractors},
Phys. Rev. \textbf{D54 }(1996) 1525, \texttt{hep-th/9603090}.

\bibitem{FGK} S.~Ferrara, G.~W.~Gibbons, R.~Kallosh, \textit{Black holes and
critical points in moduli space}, Nucl.\ Phys.\ \textbf{B500}, 75 (1997),
\texttt{hep-th/9702103}.

\bibitem{Moore:1998pn} G.~W.~Moore, \textit{Arithmetic and attractors},
\texttt{hep-th/9807087}. G.~W.~Moore, \textit{Attractors and arithmetic},
\texttt{hep-th/9807056}.

\bibitem{Kallosh:1999mz} R.~Kallosh, A.~D.~Linde, M.~Shmakova, \textit{%
Supersymmetric multiple basin attractors}, JHEP \textbf{9911}, 010 (1999),
\texttt{hep-th/9910021}.

\bibitem{Kallosh:1999mb} R.~Kallosh, \textit{Multivalued entropy of
supersymmetric black holes}, JHEP \textbf{0001}, 001 (2000), hep-th/9912053.

\bibitem{Win} M. Wijnholt, S. Zhukov, \textit{On the uniqueness of black
hole attractors}, \texttt{hep-th/9912002}.

\bibitem{Ceresole:2007rq} A.~Ceresole, S.~Ferrara, A.~Marrani, \textit{4d/5d
Correspondence for the Black Hole Potential and its Critical Points},
Class.\ Quant.\ Grav.\ \textbf{24}, 5651 (2007), \texttt{arXiv:0707.0964
[hep-th]}.

\bibitem{Dominic:2014zia} P.~Dominic, T.~Mandal, P.~K.~Tripathy, \textit{%
Multiple Single-Centered Attractors}, JHEP \textbf{1412}, 158 (2014),
\texttt{arXiv:1406.7147 [hep-th]}.

\bibitem{Manda:2015zoa} T.~Mandal, P.~K.~Tripathy, \textit{On the Uniqueness
of Supersymmetric Attractors}, Phys.\ Lett.\ \textbf{B749}, 221 (2015),
\texttt{arXiv:1506.06276 [hep-th]}.

\bibitem{Tripathy-nBPS} P.~K.~Tripathy, \textit{New Branches of
Non-supersymmetric Attractors in }$\mathcal{N}\mathit{=2}$\textit{\
Supergravity}, \texttt{arXiv:1701.00368 [hep-th]}.

\bibitem{Cremmer:1979up} E.~Cremmer and B.~Julia, \textit{The }$\mathit{SO(8)%
}$\textit{\ Supergravity}, Nucl.\ Phys.\ \textbf{B159}, 141 (1979).

\bibitem{Hull:1994ys} C.~M.~Hull and P.~K.~Townsend, \textit{Unity of
superstring dualities},\ Nucl.\ Phys.\ \textbf{B438}, 109 (1995), \texttt{%
hep-th/9410167}.

\bibitem{Borsten} L. Borsten, D. Dahanayake, M. J. Duff, W. Rubens, \textit{%
Black holes admitting a Freudenthal dual}, Phys. Rev. \textbf{D80}, 026003
(2009), \texttt{arXiv:0903.5517 [hep-th]}.

\bibitem{FMY-F} S. Ferrara, A. Marrani, A. Yeranyan, \textit{Freudenthal
duality and generalized special geometry}, Phys. Lett. \textbf{B701}, 640
(2011), \texttt{arXiv:1102.4857 [hep-th]}.

\bibitem{DualitiesNH} S. Ferrara, A. Marrani, E. Orazi, M. Trigiante, \textit{Dualities Near the Horizon}, JHEP \textbf{1311}, 056
(2013), \texttt{arXiv:1305.2057 [hep-th]}.

\bibitem{Marrani} A. Marrani, \textit{Freudenthal Duality in Gravity: from Groups of Type $E_7$ to Pre-Homogeneous Spaces}, p Adic Ultra.Anal.Appl. \textbf{7}, 322 (2015), \texttt{arXiv:1509.01031 [hep-th]}.

\bibitem{FG} A. Marrani, C. X. Qiu, S. Y. D. Shih, A. Tagliaferro, B.
Zumino, \textit{Freudenthal gauge theory}, JHEP \textbf{1303}, 132 (2013),
\texttt{arXiv:1208.0013 [hep-th]}.

\bibitem{FM} J. J. Fern\'{a}ndez-Melgarejo, E. Torrente-Lujan, $\mathcal{N}%
\mathit{=2}$\textit{\ sugra BPS multi-center solutions, quadratic
prepotentials and Freudenthal transformations}, JHEP \textbf{1405}, 081
(2014), \texttt{arXiv:1310.4182 [hep-th]}.

\bibitem{L-F-Dual} L. Borsten, M. J. Duff, S. Ferrara, A. Marrani, \textit{%
Freudenthal dual Lagrangians}, Class. Quant. Grav. \textbf{30}, 235003
(2013), \texttt{arXiv:1212.3254 [hep-th]}.

\bibitem{Klemm} D. Klemm, A. Marrani, N. Petri, M. Rabbiosi, \textit{%
Nonlinear symmetries of black hole entropy in gauged supergravity}, \texttt{%
arXiv:1701.08536 [hep-th}].

\bibitem{Shmakova:1996nz} M.~Shmakova, \textit{Calabi-Yau black holes},
Phys.\ Rev.\ \textbf{D56}, 540 (1997), \texttt{hep-th/9612076}.

\bibitem{dWVP} B. de Wit, A. Van Proeyen, \textit{Special geometry, cubic
polynomials and homogeneous quaternionic spaces}, Commun. Math. Phys.
\textbf{149}, 307 (1992), \texttt{hep-th/9112027}. B. de Wit, A. Van
Proeyen, \textit{Isometries of special manifolds}, \texttt{hep-th/9505097}.

\bibitem{dWVVP} B. de Wit, F. Vanderseypen, A. Van Proeyen, \textit{Symmetry
structure of special geometries}, Nucl. Phys. \textbf{B400}, 463 (1993),
\texttt{hep-th/9210068}.

\bibitem{alg-geom} S. Hosono, B. H. Lian, S.-T. Yau, \textit{GKZ generalized
hypergeometric systems in mirror symmetry of Calabi-Yau hypersurfaces},
Commun. Math. Phys. \textbf{182}, 535 (1996), \texttt{alg-geom/9511001}.

\bibitem{DF} R. D'Auria, S. Ferrara, \textit{String quantum symmetries from
Picard-Fuchs equations and their monodromy}, Annals Phys. \textbf{231}, 84
(1994).

\bibitem{d-geom-revisited} A. Ceresole, S. Ferrara, A. Gnecchi, A. Marrani, $%
\mathit{d}$\textit{-Geometries Revisited}, JHEP \textbf{1302}, 059 (2013),
\texttt{arXiv:1210.5983 [hep-th]}.

\bibitem{Klemm-quantum-STU} D. Klemm, A. Marrani, N. Petri, C. Santoli,
\textit{BPS black holes in a non-homogeneous deformation of the stu model of
}$\mathcal{N}\mathit{=2}$\textit{, }$\mathit{D=4}$\textit{\ gauged
supergravity}, JHEP \textbf{1509}, 205 (2015), \texttt{arXiv:1507.05553
[hep-th]}.

\bibitem{Ferrara:2006yb} S.~Ferrara, E.~G.~Gimon, R.~Kallosh, \textit{Magic
supergravities, }$\mathcal{N}\mathit{=8}$\textit{\ and black hole composites}%
, Phys.\ Rev.\ \textbf{D74}, 125018 (2006), \texttt{hep-th/0606211}.

\bibitem{Cortes} V. Cort\'{e}s, M. Dyckmanns, M. J\"{u}ngling, D. Lindemann,
\textit{A class of cubic hypersurfaces and quaternionic K\"{a}hler manifolds
of co-homogeneity one}, \texttt{arXiv:1701.07882 [math.DG]}.
\end{thebibliography}
\end{document}